\documentclass[aps,prl,reprint,nofootinbib]{revtex4-1}
\usepackage{blindtext}
\usepackage{graphicx}  
\usepackage{dcolumn}  
\usepackage{bm}        
\usepackage{amssymb}
\usepackage{amsmath,mathtools,cases, empheq} 
\usepackage{epstopdf} 
\usepackage{amsfonts,float}
\usepackage{fancyhdr}
\usepackage{pifont}

\usepackage[hidelinks]{hyperref} 
\usepackage{bbm}
\usepackage{tensor}
\usepackage{mathrsfs}
\usepackage{xcolor}
\usepackage{physics}

\usepackage[titles]{tocloft}
\usepackage{eqparbox}

		\newcommand{\bea}{\begin{eqnarray}} 
		\newcommand{\eea}{\end{eqnarray}}

\newcommand{\qed}{\nobreak \ifvmode \relax \else
      \ifdim\lastskip<1.5em \hskip-\lastskip
      \hskip1.5em plus0em minus0.5em \fi \nobreak
      \vrule height0.75em width0.5em depth0.25em\fi}

\begin{document}
\title{Casimir effect and \textit{TGTG}-formula for curved backgrounds}
\author{Luc{\'i}a Santamar{\'i}a-Sanz}
\email{lucia.santamaria@uva.es}
\affiliation{Departamento de F{\'i}sica Te{\'o}rica, At{\'o}mica y {\'O}ptica, Universidad de Valladolid, 47011 Valladolid, Spain}

\begin{abstract}
The quantum vacuum interaction energy between a pair of semitransparent two-dimensional plates in the topological background of a sine-Gordon kink is studied. Quantum vacuum oscillations around the sine-Gordon kink solution can be interpreted as a quantum scalar field theory in the spacetime of a domain wall. An extension of the \textit{TGTG}-formula, firstly discovered by O. Kenneth and I. Klich, to weak curved backgrounds is obtained. 
\end{abstract}

\maketitle

\section{Introduction}
Topological solitons  are finite energy solutions of classical field equations that asymptotically connect two nonequivalent vacua in different topological sectors of the configurational space. Therefore, they break discrete symmetries \cite{Colemanbook, Rajaramanbook, Vachaspatibook}. In particular, topological solitons and their quantization are used as models for extended fundamental objects such as domain walls  \cite{Vachaspatibook}.  Quantum vacuum fluctuations of scalar fields $\phi(x^\mu)$ around these classical solutions and their interaction with the background have been extensively studied in Quantum Field Theory (QFT)  \cite{Mantonbook}. One well-known example is the 1+1 dimensional sine-Gordon model, described by the Lagrangian density
\begin{equation*}
\mathcal{L}= \frac{1}{2}\partial_\mu \phi \partial^\mu \phi +\cos\phi -1, \qquad  x^\mu=(t,z),
\end{equation*}
in terms of  dimensionless magnitudes. When studying small fluctuations $\eta(t,z)$ around the classical static solution  $\phi(z)= 4 \arctan e^z$ up to second order, a Lagrangian quadratic in the fluctuations is obtained
\begin{equation*}
\mathcal{L}= \frac{1}{2}\left[\partial_\mu \eta \partial^\mu \eta -\left(1-2 \sech^2 z\right) \eta^2\right].
\end{equation*}
The propagation of mesons around this background will be the starting point of this work.

The finite part of the one-loop self-interaction vacuum diagrams added all together is the so-called zero point energy \cite{Miltonbook}. One of its most known manifestation in nature is the Casimir effect, firstly discovered by H.B.G. Casimir in 1948 \cite{Casimir1948}, and measured in the laboratory by M.J. Sparnaay \cite{Sparnaay1957}. The Casimir force appears as a result of the difference between the energy of the fluctuating field when static or slowly changing external  objects are introduced in the space and when they are removed. These forces have been studied in \cite{Bordagbook} for bodies of different shapes and compositions. They have numerous applications in nanoelectronic devices \cite{Fosco2009, Allen2005}, in absorption phenomena in carbon nanotubes \cite{Klimchitskaya2008} and in inflation process \cite{Mamaevbook}, to name just a few of examples.

The direct calculation of the quantum vacuum energy as the expectation value of the field theoretical Hamiltonian for the vacuum state ($E_{vac}\sim\langle 0\vert{\bf H}\vert 0\rangle$) gives an infinite answer. It is difficult to remove the inherent ultraviolet divergences that appears because of the infinite number of fluctuating modes present in the problem. Many methods to renormalize these divergences have been studied \cite{Bordagbook}. One of the most important theoretical results is the so-called \textit{TGTG}-formula, firstly introduced by O. Kenneth and I. Klich \cite{KK2008} for studying two bodies separated by a small distance in flat spacetimes. It is a formalism based on transition operators (more precisely the so called Lippmann-Schwinger \textit{T}-operator) and Green's functions. The \textit{TGTG}-formula gives the distance-dependent finite part of the zero-point energy in terms of the non-relativistic quantum scattering data for each of the objects. It is exact whenever the bodies are represented by potentials with disjoint compact supports. 

When considering curved spacetimes, the solutions of the field equations, the temporal coordinate and the usual canonical quantization are globally defined only for globally hiperbolic spacetimes endowed with a global temporal Killing vector \cite{Fewster2008}. Whenever there is a foliation in Cauchy surfaces, one could solve the spectra of the Laplacian-Beltrami operator in the spatial slice for each fixed value of the temporal coordinate as done in Minkowski metrics. However, in more general cases, if the curved spacetime is such that the fiber bundle do not allow an interpretation in terms of particle spectra independent of the observer, neither scattering nor transfer operators  will be universal results. In fact, there are not many results about the \textit{TGTG}-formula in curved backgrounds and sometimes it does not even exist \cite{BordagJMMC2014}. However, there is a special case to be taken into account. When the frequencies of the particles created by the gravitational background are much smaller than the Planck frequency, one could use the perturbation theory as a semiclassical approach to quantum gravity \cite{Bar2009, Kay2006}.  In doing so, this weak gravitational backgrounds are treated classically and the matter fields are the ones which will be quantized.  Here, an example of weak gravitational field is going to be studied: a 3+1 dimensional spacetime with a P\"oschl-Teller (PT) kink, $V_{PT}(z)=-2/\cosh^2(z)$, in one of the spatial dimensions (notice that $x^\mu=(t, \vec{x}_\parallel, z), \vec{x}_\parallel\in \mathbb{R}^2$). This gravity would be strong enough to produce some effects to the quantum matter but not so strong as to require an own quantization. Furthermore, the PT potential is transparent in the sense that the fields could be asymptotically interpreted as particles. Thus it is possible to define incoming and outgoing waves and to derive a $S$-matrix in a similar way to the usual for flat spacetimes. Furthermore, since $V_{PT}$ is transparent, there is not additional reflection with respect to the free case.

The aim of this work is to derive a \textit{TGTG}-formula for studying the quantum vacuum interaction energy between two homogeneous plates, mimicked by a point potential  $V_{2\delta}(z)=v_0\delta(z-a) + v_1 \delta(z-b), \,\, v_0, v_1 \in \mathbb{R}$,  embedded in the PT potential background $V_{PT}$ centered at $z=0$. In the last two decades, potentials supported on a point, or equivalently contact interactions \cite{Gadella2009}, have been used to mimic plates and other geometries in the Casimir setup (see  \cite{Hennig1992, Guilarte2013} and references therein). In fact, the electromagnetic properties of conducting plates can be studied through delta interaction couplings, which represent the plasma frequencies in Barton's hydrodynamical models \cite{Barton2004}.

The present work is organized as follows. Firstly, the quantum mechanical spectrum of two  Dirac $\delta$ interactions in the curved background of the kink is described.  Then, the quantum interaction between  plates is studied through the Green's function and the $T$-operator in the \textit{TGTG}-formalism. Finally, some conclusions are collected. The natural system of units $\hbar=c=1$ will be used throughout the text. The full technical details and additional developments are deferred to the companion paper \cite{Santamariasupl}.

\section{Scattering data and spectrum}
The Casimir force between plates is due to the coupling between the quantum vacuum fluctuations of the electromagnetic field with the charged current fluctuations of the plates \cite{Lifshitz1956}. For distances between plates rather larger than any other length scale concerning the electric response of the plates, only the long wavelength transverse modes of the electromagnetic field are relevant to the interaction. They can be mimicked by the normal modes of a scalar field whose dynamics is described by the action
\begin{eqnarray*}
S[\phi] = \frac{1}{2} \int d^4x \left[\partial_\mu \phi \partial^\mu \phi -V_{2\delta}(z)\phi^2-V_{PT} (z) \phi^2\right],
\end{eqnarray*}
where $v_0,v_1, a< b \in \mathbb{R}$. Notice that $z$ would be the coordinate of the dimension orthogonal to the plates and $x \equiv x^\mu$. I am also considering that the temperature of the system is zero. A detailed description of the spectrum of the associated non-relativistic Schr\"odinger operator $-\partial^2_{\vec{x}_\parallel}+\hat{K}= -\partial^2_{\vec{x}_\parallel} -\partial_z^2 +V_{PT}(z)+V_{2\delta}(z)$  is needed to identify the eigenmodes of the scalar field fluctuations.  The system has an open geometry so the positive energy spectrum would be continuous. Scattering states correspond to solutions of the Schr\"odinger equation $\hat{K} \phi = k^2 \phi$ with $k\in\mathbb{R}$ (such that $k^2 > 0$). Away from the singular points, the ``\textit{diestro}" scattering solutions (incoming particles from the left) are of the form:
\begin{eqnarray}\label{soldiestro}
\!\!\!\!\psi^R_k(z)= \left\{ \begin{array}{lll}
              f_k(z)+r_R \, f_{-k}(z) , && \textrm{if} \, \,\,  z<a,\\[1ex]
              B_R\,  f_k(z) +C_R\, f_{-k}(z) , && \textrm{if} \,\,\,   a<z<b, \\[1ex]
              t_R \, f_k(z),  && \textrm{if} \,\,\, z>b,
             \end{array}
   \right. 
\end{eqnarray}
being $f_k(z)=e^{ikz}(\tanh (z)-ik)$ the  free waves of the P\"oschl-Teller potential. Likewise, the wave function of ``\textit{zurdo}" scattering (incoming particles from the right) can be described as:
\begin{eqnarray}\label{solzurdo}
\!\!\!\!\psi^L_k(z)= \left\{ \begin{array}{lll}
            t_L\, f_{-k}(z), && \textrm{if} \, \,\,  z<a,\\[1ex]
              B_L\,  f_{-k}(z)+C_L\,  f_{k}(z), && \textrm{if} \, \,\,  a<z<b, \\[1ex]
             r_L \, f_k(z) +  f_{-k}(z), && \textrm{if} \, \,\, z>b.
             \end{array}
   \right. 
\end{eqnarray}
$\{t_L, t_R, r_L, r_R, B_L, B_R, C_L, C_R\}(k)$ are the scattering data. The transmission amplitudes verify $t_R(k)=t_L(k)$ due to the time-reversal invariance of the Schr\"odinger operator. From now on, they will be replaced by $t(k)$.

Notice that the operator $\hat{K}^{PT}_z=-\partial_z^2+V_{PT}(z)$ is not essentially self-adjoint when its domain is the Sobolev space of functions $W^2_2(\mathbb{R} -\{a,b\}, \mathbb{C})$.  For defining the self-adjoint extensions it is necessary to add some matching conditions at the boundary points $\{a,b\}$:
\begin{eqnarray}\label{kurasov}
\hspace{-0.8cm}\left(
\begin{array}{c}
\psi(a^+)\\
\psi^{\prime}(a^+)\\
\psi(b^+)\\
\psi^{\prime}(b^+)
\end{array}\right)=
\left(
\begin{array}{cccc}
1 & 0 &0 &0\\[0.5ex]
 v_0 & 1&0 &0\\
 0 &0 & 1 & 0 \\[0.5ex]
0 &0 &  v_1 & 1
\end{array}
\right)\left(
\begin{array}{c}
\psi(a^-)\\
\psi^{\prime}(a^-)\\
\psi(b^-)\\
\psi^{\prime}(b^-)
\end{array}\right).
\end{eqnarray}
Above, $\psi(z^\pm)$ means the right-hand ($+$) or left-hand ($-$) limit of the wave function at the point $z$. 

Replacing \eqref{soldiestro} and \eqref{solzurdo} in \eqref{kurasov} and solving the resulting two systems of equations, one obtains the scattering data. The transmission and reflection coefficients are collected in \eqref{scatteringkink}, where $W=W[f_{k}(x), f_{-k}(x)]= -2 i k (k^2+1)$. Notice that the scattering data explicitly depend on the position of the plates in a non-trivial way because the P\"oschl-Teller background potential breaks the translational invariance in the space.

The common denominator of all the scattering parameters is the spectral function.  Its  zeroes on the positive imaginary axis gives the bound states of the spectrum of $\hat{K}$. It can be checked that there are no bound states with energy below a certain quantity $E_{min}$, which takes the value:
\begin{eqnarray}\label{Emin}
\!\! \!\!  E_{min}=-\frac{1}{16}\left[(|v_1|+|v_0|)+\sqrt{(|v_1|+|v_0|)^2+16}\right]^2\! ,
\end{eqnarray}
for all $a,b,v_0,v_1 \neq 0$.  
\begin{widetext}
\begin{eqnarray}\label{scatteringkink}
t&=& \frac{W^2}{W^2 + v_0 W f_{k}(a)f_{-k}(a)+v_1 W f_{k}(b)f_{-k}(b) +v_0 v_1 f_{-k}(a)f_{k}(b) [f_{k}(a)f_{-k}(b)-f_{-k}(a)f_{k}(b)]},\nonumber\\
r_R&=& \frac{-v_0 W f_{k}^2(a)-v_1 W f_{k}^2(b)+v_0 v_1 f_{k}(a)f_{k}(b) [f_{-k}(a)f_{k}(b)-f_{k}(a)f_{-k}(b)]}{W^2 + v_0 W f_{k}(a)f_{-k}(a)+v_1 W f_{k}(b)f_{-k}(b) +v_0 v_1 f_{-k}(a)f_{k}(b) [f_{k}(a)f_{-k}(b)-f_{-k}(a)f_{k}(b)]},\\
r_L&=& \frac{-v_0 W f_{-k}^2(a)-v_1 W f_{-k}^2(b)-v_0 v_1 f_{-k}(a)f_{-k}(b) [f_{k}(a)f_{-k}(b)-f_{-k}(a)f_{k}(b)]}{W^2 + v_0 W f_{k}(a)f_{-k}(a)+v_1 W f_{k}(b)f_{-k}(b) +v_0 v_1 f_{-k}(a)f_{k}(b) [f_{k}(a)f_{-k}(b)-f_{-k}(a)f_{k}(b)]}.\nonumber
\end{eqnarray}
\end{widetext}

The value of $E_{min}$ is essential for the computation of the quantum vacuum interaction energy in the corresponding QFT. Since the bound state with the lowest energy is characterized by $E_{min}$, the mass of the fluctuations in the theory will be balanced with this value  $E_{min}$ for making fluctuation absorption impossible. The unitarity of the QFT sets this lower bound for the mass of the quantum vacuum fluctuations, so that the total energy of the lowest energy state of the spectrum will be zero.   In this way, the spectrum will be formed by a set of discrete states with energies within the gap $[0, m^2]$ and a continuum of scattering states with energies above this threshold, i.e. $E=m^2=|E_{min}|$.

\section{\textit{TGTG}-formula and Casimir pressure}
After second quantization, the non-relativistic quantum mechanical problem explained in the previous section is promoted to a QFT.  The objective now is to study the quantum vacuum energy $\tilde{E_0}=\frac{1}{2} \sum_{\omega^2} \omega $, where the summation is performed over the spectrum of the operator $-\partial_{\vec{x_\parallel}}^2+\hat{K}^{PT}_z+V_{2\delta}(z)$. The modes of the quantum field satisfy $\omega^2=\vec{k}_\parallel^2+k^2+m^2$. Notice that $k_\parallel$ refers to the component of the momentum in the two directions parallel to the surfaces of the plates and $k$ to the orthogonal one. In order to compute the interaction energy, $\tilde{E_0}$ must be regularized and renormalized to remove the ultraviolet divergences due to the infinite energy density of the field theory in the bulk and the subdominant divergences associated to the infinite area of the plates. Firstly, a regulator parameter is introduced in the expression of the vacuum energy per unit area of the plates
\begin{equation*}
\frac{E_0}{A}= \lim_{\epsilon \to 0}\frac{1}{2} \int_{\mathbb{R}^2} \frac{d\vec{k}_\parallel}{(2\pi)^2}\, \,  \underset{k}{\mathclap{\displaystyle\int}\mathclap{\textstyle\sum}}   \, \, \sqrt{\vec{k}_\parallel^2+k^2+m^2} \,\, e^{-\epsilon (\vec{k}_\parallel^2+k^2+m^2)}
\end{equation*}
and then, the divergent terms arising in the expansion of the integrand around $\epsilon\to 0$ are removed. 

The \textit{TGTG}-formula for flat spacetimes \cite{KK2008} reads
\begin{equation}\label{TGTGflat}
\!\!\!\!E_0=-i \! \int_0 ^\infty \!\! \frac{d\omega}{2\pi} \log \!\left[1-\tr (\!T^{\ell}G^{(0)}\!(a,b)T^rG^{(0)}\!(b,a)\!)\!\right],
\end{equation}
with $G^{(0)}$ the Green's function of the plain background, without plates. The superscript $\ell, r$ indicates which plate is being considered: $\ell$ for the plate placed on the left on the system and $r$ for the other one. There are two fundamental advantages in this formalism. On the one hand,  when using the scattering data to compute the Casimir energy between several objects, it is necessary to characterize the whole spectrum in the background of these two objects, which can frequently be rather difficult. By contrast, since the \textit{TGTG}-formula only requires knowing the spectrum of a single object, the computational effort is much smaller. On the other hand, the \textit{TGTG}-formula is free of divergences and directly gives the Casimir energy between two separated bodies. Consequently,  this formalism is a convenient option among all existing regularization procedures.

Since I deal with a weak and transparent background potential, it is reasonable to look for a generalization of the \textit{TGTG}-formula that applies to this specific type of curved manifolds. This purpose requires computing the Green's functions and the transfer operator related to each plate.

The Green's function solution of the equation
\begin{eqnarray*}
\!\left[\partial_t^2-\partial_{\vec{x}_\parallel}^2\! + \hat{K}^{PT}\!+m^2+ V_{2\delta}(z)\right]\! G(x^\mu, y^\mu)=\delta(x^\mu-y^\mu),
\end{eqnarray*}
can be expressed as
\begin{equation*}
G(x,x') = \int \frac{d^{2}k_\parallel}{(2\pi)^{2}} e^{i \vec{k}_\parallel (\vec{x}_\parallel-\vec{x'}_\parallel)} \int \frac{d\omega}{2\pi} e^{-i \omega (t-t')}G_k(z,z').
\end{equation*}
The reduced Green's function on the direction orthogonal to the plates can be computed as
\begin{equation*}
 G_k(z,z') \!= \!\frac{\textrm{u}(z-z')\psi_k^R(z)\psi_k^L(z')\!+\! \textrm{u}(z'-z)\psi_k^R(z')\psi_k^L(z)}{W[\psi_k^R, \psi_k^L]} 
\end{equation*}
from the scattering solutions \eqref{soldiestro}-\eqref{solzurdo} and the unit step function. The result of this calculation can be written as $G_{k}(z_1, z_2)= G^{PT}_k (z_1, z_2) +\Delta G_{k}(z_1, z_2)$. Notice that $G^{PT}_k (z_1, z_2) = f_{-k}(z_<)f_{k}(z_>)/W$ (being $z_<$ and $z_>$ the lesser or the greater of $\{z_1, z_2\}$) is the Green's function for the PT potential without any delta interactions. Moreover,
\begin{empheq}[left= {\hspace{-5pt} \Delta G_{k}(z_1, z_2)= \empheqlbrace}]{align}
&\begin{aligned}
  & \frac{r_L}{W} f_{k}(z_1)f_{k}(z_2), & &\text{if $z_1, z_2>b$,} \\
  &\frac{r_R}{W} f_{-k}(z_1)f_{-k}(z_2), & &\text{if $z_1, z_2<b$},\\
  &(t-1) \, G^{PT}_k (z_1, z_2), & &\text{otherwise},
\end{aligned}
\label{greenoneplate}
\end{empheq}
is the remaining part of the Green's function when one of the two plates is removed (i.e $v_0=0$). When both plates are introduced into the system, a more complicated expression for $\Delta G_{k}(z_1, z_2)$ is obtained but it will not be included here because it will not be necessary for the calculation of the \textit{TGTG}-formula. 

From the Lippmann-Schwinger equation and the definition of Green's function, it is possible to compute the transfer operator of the plate placed at $z=b$ as
\begin{eqnarray*}
T_k(z_1, z_2) &=& - (t-1) (\hat{K}_{z_2}^{PT}-k^2)  \, (\hat{K}_{z_1}^{PT}-k^2)  \, G^{PT}_k (z_1, z_2)\nonumber\\&=&\frac{|W|^2}{k^4} \delta(z_1-b) \delta(z_2-b)  \Delta G_k(b,b).
\end{eqnarray*}
Here $\Delta G_k(b,b)$ is given by \eqref{greenoneplate}. The normalization of the \textit{T}-operator due to its relation to the scattering matrix yields
\begin{eqnarray}\label{toperator}
&&\!T(z_1, z_2) \!=\!- W^* \delta(z_1\! -b) \delta(z_2\! -b)\!\left\{\begin{array}{lc}
\! r_L(b),& \!\! \!\text{$z_1, z_2 \to b^+$\!,} \\
\! r_R(b), & \!\! \!\text{$z_1, z_2 \to b^-$}\!,\\
\! 1-t(b),  & \!\!\! \text{otherwise},
\end{array}\right.\nonumber\\&&
\end{eqnarray}
and analogously for the plate on the left. Asterisk means complex conjugate. This transfer operator represents the probability amplitude for a particle to interact with the potential but without propagation.

Combining \eqref{greenoneplate}, \eqref{toperator} and generalizing \eqref{TGTGflat} to three spatial dimensions before performing a Wick rotation in the momentum, yields the version of the \textit{TGTG}-formula for the weak transparent curved spacetime considered:

\vspace{-10pt}{\small \begin{eqnarray}\label{TGTGcurved}
\frac{E_0}{A}\! &=&\! \frac{1}{8\pi^2} \! \int_{m}^{\infty} \!\!\! \! d\xi\,  \xi\,  \sqrt{\xi^2-m^2} \, \textrm{log} \!\left(\! 1\! -\textrm{Tr}\! \left. \left[T^\ell G^{PT}T^r G^{PT}\right]\right|_{i\xi}\right)\nonumber\\
 &-&\frac{1}{2}\sum_{j=1}^N\frac{\left(\sqrt{(i\kappa_j)^2+m^2}\right)^{3}}{6\pi},
\end{eqnarray} }
being
\begin{eqnarray*}
\textrm{Tr}\, \textrm{TGTG}\biggr\vert_{k} =r_R^r(k,v_1,b) \, r_L^\ell(k,v_0,a). 
\end{eqnarray*}
 Notice that $r_R^r, r_L^\ell$ can be obtained from \eqref{scatteringkink} by replacing $v_0=0$ and $v_1=0$, respectively. The most interesting detail about this result is that the only combinations of the \textit{T}-operator components that allow coincidences of the $z_1, z_2$ points within the interval $[a,b]$, and thus contribute to the quantum vacuum interaction energy between plates, depend only on these two reflection coefficients. This is also the case for flat spacetimes \cite{KK2008}. But for curved backgrounds that breaks the isotropy of the space, the scattering data depend explicitly on the position of each plate, and either the propagator and the characteristic waves of the specific background have to be taken into account. Furthermore, the integral \eqref{TGTGcurved} is well define because the \textit{TGTG}-operator is a trace class one for configurations with separated bodies and the modulus of its eigenvalues are smaller than one (further details in \cite{KK2008}).

\begin{figure}[H]
\centering
 \includegraphics[width=0.49\textwidth]{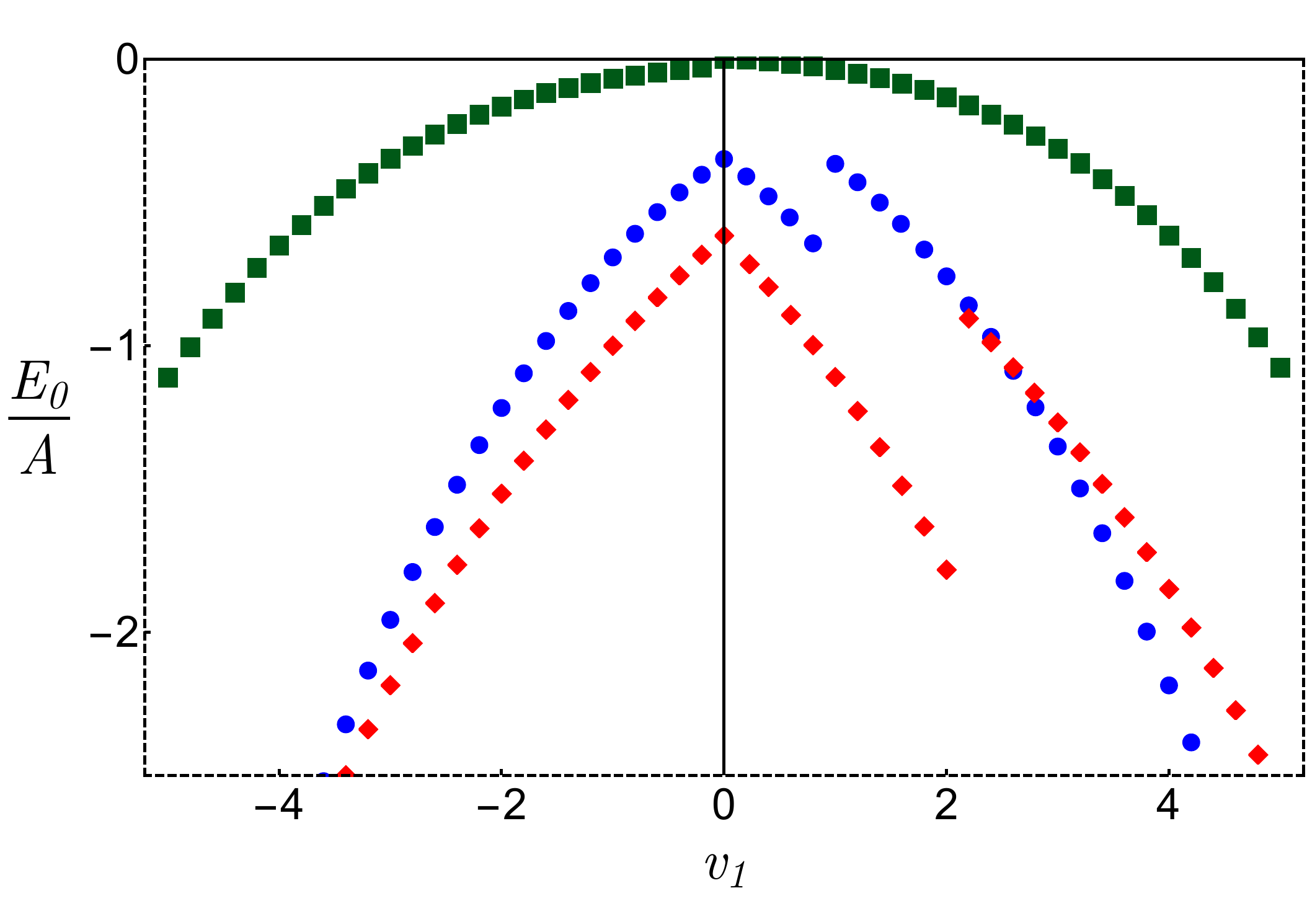}
\caption{\footnotesize $E_0$ per unit area between plates for $a=-b=-0.5$ as a function of $v_1$. In this plot $v_0=-3$ (circles), $v_0=0.1$ (squares) and $v_0=4$ (rhombi).}
\label{fig:PTdentro}
\end{figure}

Figure \ref{fig:PTdentro} shows the quantum vacuum interaction energy for a symmetrical plate configuration in the PT background. As can be seen, the energy is always negative, independently of the value of the coefficients of the delta potentials. This implies that the Casimir force between plates will always be attractive in this system. If the P\"oschl-Teller well was not confined at all between the plates and they were far from the kink center, the system of two $\delta$-plates in flat spacetime would be recovered. 

The sudden jump discontinuities present in FIG. \ref{fig:PTdentro} are related to the loss of one bound state with very low momentum in the spectrum of the fluctuation field. It can also be seen that the larger the magnitude of the delta coefficients, the larger the one of $E_0$.  Finally, it is worth pointing that even in the case where one of the delta coefficients is zero (and as a consequence, there is only one plate in the system), the other plate feels the interaction since there is still a non-zero Casimir energy in the system.

\section{Conclusions}
In this work, a quantum scalar field interacting with two parallel two-dimensional plates mimicked by Dirac $\delta$-potentials in a curved background of a topological P\"oschl-Teller kink at zero temperature is presented. The quantum vacuum interaction energy has been calculated through a generalization of the \textit{TGTG}-formula for weak and transparent gravitational backgrounds.  The result only depends on the reflection coefficients associated to the scattering problem. The special features of the curved background chosen are implicit in the definition of these scattering coefficients. The virtue of the \textit{TGTG}-formula obtained here is that it can be easily generalized to other type of configurations, either for another background and for other potentials that could properly mimic the plates.  In fact, there are evidence \cite{Romaniega2021} that the introduction of the first derivative of the delta potential for mimicking spherical shells causes the sign of the Casimir force between them to change in different areas of the parameter space. This type of $\delta \delta'$ plates will be studied in the supplementary material \cite{Santamariasupl}. 

I would like to finish by mentioning a possible future phenomenological application of the study collected in this work. Notice that in the Casimir effect, the electromagnetic force results from the computation of the one loop quantum correction to the vacuum polarization in QED. Nevertheless, taking a quadratic approximation for the action implies that it does not depend on the coupling constant of the theory, and for that reason the force is relevant at the nanometre scale. By the same argument, if one loop quantum corrections to the graviton propagator were calculated, the coupling constant of the gravitational theory would not appear either. And therefore, since the graviton has zero mass, one could think of studying the quantum interacting force between gravitational objects described by sufficiently strong fields separated by a small distance, using the same quadratic approximation reasoning applied here. The force caused by non-massive quantum vacuum fluctuations around a classical solution for gravitons coupled to no matter which other massive field would constitute an example of quantum corrections to the gravitational field theory. The difficulty will lie in finding a material that is opaque to the gravitational waves \cite{Quach2015}.

\section{Acknowledgments}
I am grateful to C. Romaniega, LM. Nieto, I. Cavero-Pel\'aez and JM. Mu\~{n}oz-Casta\~{n}eda for fruitful discussions. This research was supported by Spanish MCIN with funding from European Union NextGenerationEU (PRTRC17.I1) and Consejer\'ia de Educaci\'on from JCyL through QCAYLE project, as well as MCIN project PID2020-113406GB-I00.  I am thankful to the Spanish Government for the FPU PhD fellowship program (Grant No. FPU18/00957).

\bibliographystyle{apsrev4-1}
\bibliography{biblio}  

\end{document}